\documentclass[12pt,preprint]{aastex}

\shorttitle{The Librating Companions: Alignment or Antialignment?}
\shortauthors{Ji Jianghui et al.}

\begin{document}

\title{The Librating Companions in HD 37124, HD 12661, HD 82943,
47 Uma and GJ 876: Alignment or Antialignment?}

\author{Jianghui JI\altaffilmark{1,2}, Lin
LIU\altaffilmark{3}, Hiroshi Kinoshita\altaffilmark{4}, Jilin
ZHOU\altaffilmark{3}, Hiroshi Nakai\altaffilmark{4}, Guangyu
LI\altaffilmark{1,2}} \email{jijh@pmo.ac.cn }

\altaffiltext{1}{Purple  Mountain  Observatory , Chinese  Academy
of  Sciences ,  Nanjing  210008,China}

\altaffiltext{2}{National Astronomical Observatory, Chinese
Academy of Sciences,Beijing 100012,China}

\altaffiltext{3}{Department of Astronomy,  Nanjing University,
Nanjing  210093, China}

\altaffiltext{4}{National Astronomical Observatory,
 Mitaka, Tokyo 181-8588,Japan}

\begin{abstract}
We investigated the apsidal motion for the multi-planet systems.
In the simulations, we found that the two planets of HD 37124, HD
12661, 47 Uma and HD 82943 separately undergo apsidal alignment or
antialignment. But the companions of GJ 876 and $\upsilon$ And are
only in apsidal lock about $0^{\circ}$. Moreover, we obtained the
criteria with Laplace-Lagrange secular theory to discern whether a
pair of planets for a certain system are in libration or
circulation.
\end{abstract}

\section{Introduction}
The number of known extrasolar planetary systems  are quickly
growing (Butler et al. 2003,Paper I) in recent years. The analysis
and orbital fitting of the curves of radial velocity can reveal
the fact that the main-sequence stars may host one or more
planets. Recently, Fischer et al.(2003)(Paper II) reported the
properties of ten multiple-planet systems (see Table 8 in Paper
II) and pointed out that most of the systems with the ratio of
orbital periods less than 5:1 are characterized by mean motion
resonance (MMR), such as GJ 876(2:1) and HD 82943(2:1). The other
important feature for these systems is the apsidal lock of the
orbiting companions, which indicates that the relative apsidal
longitudes of two orbits librate about a constant, such that two
planets have common time-averaged rate of apsidal precession. A
number of researchers studied this dynamical mechanism in the
planetary systems. The work of Ji et al.(2003a) confirmed that the
inner two companions of 55 Cnc (Marcy et al. 2002) are in a 3:1
resonance and still experience the apsidal phase-locking. In fact,
one may be familiar to the presence of the apsidal lock in our
solar system, such as the Galilean satellites of Io-Europa where
the difference of their apsidal longitudes $\Delta\varpi$ are
anti-aligned (Lee \& Peale 2002). In the case of the pulsar PSR
1257+12, Wolszczan \& Frail (1992) pointed out that there exists
almost a $\Delta\varpi$-libration about $180^{0}$ for the two
planets in a 3:2 MMR. Then a question arises: are all discovered
multiple-planet systems in apsidal lock? If so, how frequently
does this mechanism take place for these systems?

According to Laplace-Lagrange (L-L) linear secular theory (Murray
\& Dermott 1999, MD99), if $\Delta\varpi$ is a libration point,
then $360^{\circ}-\Delta\varpi$ could also be the other
equilibrium due to the symmetry of planetary configurations (also
see Beauge et al. 2003). So the second question is that if two
planets undergo alignment or antialignment, could the mirror
geometry occur? In this Letter, our goal is to examine the apsidal
motion for HD 37124, HD 12661, HD 82943, 47 Uma, GJ 876.

\section{Results}
In the present work, we performed a preliminary study of various
systems by using N-body codes(Ji, Li \& Liu 2002). We take the
masses of the parent star and companions(with sin\textit{i} = 1)
from Table 1 for each system. The time step is adopted to be 1\%
to 2.5\% of the orbital period of the innermost planet. Here we
simply account for coplanar configurations for these systems. We
remain the values of the semi-major axes unchanged for all orbits.
The eccentricities and arguments of periapse are generated in the
orbital parameters web based on the best-fit orbital solutions
given the nominal observation errors (see Table 1)\footnote{Here
we study the possible motion near the best-fit orbital solutions.
This may reveal some important dynamical features for a certain
investigated system.}. And the remaining arguments of nodal
longitudes and mean anomalies are randomly made between
$0^{\circ}$ and $360^{\circ}$. Finally,five two-planet systems
were individually produced 100 pairs of coplanar orbits for
integration from 1 Myr to 10 Myr.

Throughout the paper, $\lambda$ and $\varpi$ denote the mean and
periastron longitudes of a planet respectively. The subscripts $b$
and $c$ separately represent the inner and outer
companions\footnote{For the sake of convenience, we exchange the
order of the subscripts for two planets of HD 82943 such that the
inner planet can be always described by the symbol $b$.}.
Consequently, the relative apsidal longitude is $\Delta\varpi$ =
$\varpi_{b} - \varpi_{c}$.

\subsection{HD 37124}
Paper I revealed that two companions orbit HD 37124, but indicated
that with only 30 observations the eccentricity of the outer
planet was poorly determined,with any eccentricity between 0.3 and
0.8 fitting the observations within measurement errors (Butler
2003, private communication). In our simulations, we let $e_{c}$
initially be in the interval [0.20, 0.60] in an attempt to find
likely dynamical constraints for $e_{c}$. However, all orbits can
survive for $10^{6}$ yr, indicating the system is extremely stable
for any initial $e_{c}$ no greater than 0.60. Moreover, we noticed
that $e_{c}$ ranges from 0 to 0.62 in secular evolution for all
cases, coinciding with the outcomes by Paper I and
Kiseleva-Eggleton et al.(2002). They pointed out that the system
self-destructed for the critical initial value of $e_{c}$ above
0.65. In addition, there are two typical secular variations of
both eccentricities for stable orbits:(a)Case of large
oscillations:$e_{b}$ is up to 0.85 and $e_{c}$ is less than
0.62.(b)Case of small modulations, both $e_{b}$ and $e_{c}$ are
less than 0.40 (see Figure 1). The dynamical origin of high
eccentricity for Case (a) may result from Kozai mechanism (Kozai
1962;Wu \& Murray 2003) that implies the coupled relationship of
the maximum eccentricity and minimum inclination.

Furthermore, it is noteworthy that 43\%\footnote{The percentage of
the aligned and antialigned cases may be affected by the
uncertainty of $e_{c}$, but imply that two planets of this system
likely tend to be aligned.} and 1\% of the stable orbits are
separately in the periapse alignment and antialignment. To the
best of our knowledge, such results are a new finding for this
system. In Figure 1, the upper and low panels separately exhibit
the variations of eccentricities and relative apsidal longitudes.
The results of L-L theory (dash lines) confirm the numerical
results (thick lines) in the amplitudes and periods of libration
for both eccentricities and $\Delta\varpi$.

\subsection{HD 12661}
The orbital fits revealed that the HD 12661 system is close to
11:2 MMR (Paper II;Gozdziewski 2003a) or 6:1 MMR (Gozdziewski \&
Maciejewski 2003b). With the best-fit data from paper II, our
simulations exhibit all the orbits are stable for 1 Myr. We
observe that 46\% and 49\% of the orbits are in
$\Delta\varpi$-libration about $0^{\circ}$ and $180^{\circ}$
respectively. Moreover, Gozdziewski(2003a) indicated that the
possibility of $\Delta\varpi$-circulation is much less likely than
$\Delta\varpi$-libration about $0^{\circ}$ or $180^{\circ}$ in the
numerical investigations. Here the two types of the apsidal
librations almost share an equal opportunity to be acted as one of
the dynamical mechanisms to remain the stability of HD 12661,
because the two planets with moderate eccentricities are always
far from each other with isolated orbits and protected from close
encounters. Still, the fact of $\Delta\varpi$-libration indicates
that one planet travels through the periapse or apoapse when the
other moves to the opposite side at the same time.

The critical arguments for 11:2 MMR $\phi_{k}$ can be written:
\begin{equation}
\label{eq0} \phi_{k}= 2\lambda _{b} - 11\lambda _{c} + k\varpi_{b}
+ (9-k)\varpi_{c}, k=0...9 .
\end{equation}
Here we define $\theta_{1} = \phi_{3} = 2\lambda _{b} - 11\lambda
_{c} + 3\varpi_{b} + 6\varpi_{c}$, $\theta_{2} = \phi_{2}
=2\lambda _{b} - 11\lambda _{c} + 2\varpi_{b} + 7\varpi_{c}$ and
$\theta_{3} = \theta_{1} - \theta_{2}= \Delta\varpi$. Most systems
in a 11:2 MMR are only in behavior of $\theta_{1}$-libration. A
particular case is exhibited in Figure 2, which not only ten
resonant arguments all librate but $\theta_{3}$ librates about
$180^{\circ}$ for 1 Myr. Using octupole-level
secular theory, Lee \& Peale (2003) found that if the ratio of the
maximum orbital angular momenta is approximately equal to a
critical value,then the $\Delta\varpi$-libration about $0^{\circ}$
or $180^{\circ}$ could happen with large amplitude variations of
the eccentricities. Additionally, we further found that the status
of $\Delta\varpi$-libration are relevant to the ratio of the
initial eccentricities and the initial relative apsidal longitudes
of two planets (see \S3).

\subsection{47 Uma}
The two companions of 47 Uma is a close analog to the duo of
Jupiter and Saturn in our solar system in the ratios of the planet
masses and orbital periods. Laughlin et al.(2002) found the system
is in $\Delta\varpi$-libration about $0^{\circ}$ with low
eccentricities of the companions. Further, Gozdziewski (2002) used
MEGNO technique to investigate in wide ranges of the orbital
parameters for the system. Using the updated parameters by Paper
II, we still found this system can experience not only a
$\Delta\varpi$-libration about $0^{\circ}$, but about
$180^{\circ}$. We note that all the experiments can last for 10
Myr. The fractions in alignment, antialignment and circulation are
respectively 20\%, 10\% and 70\%. As well known, Jupiter and
Saturn are near 5:2 MMR, but the $\Delta\varpi$ circulates at
present-day. As for 47 Uma, we did find one of the critical
arguments $3 \lambda _{b} - 8\lambda _{c} + 2\varpi_{b} +
3\varpi_{c}$ temporarily librates for hundreds of years but
$\Delta\varpi$ simultaneously circulates in the secular evolution.
However, we argue that 47 Uma may be captured into 8:3 MMR in the
past during the process of migration due to the planet-disk
interaction.

\subsection{HD 82943 and GJ 876}
The celebrated nature for HD 82943 and GJ 876 is that the ratio of
the orbital periods of two companions is respectively close to
2:1. The critical arguments for 2:1 resonance are $\theta_{1} =
\lambda _{b} - 2\lambda _{c} + \varpi_{b}$, $\theta_{2} = \lambda
_{b} - 2\lambda _{c} + \varpi_{c}$, and the relative apsidal
longitude is $\theta_{3} = \theta_{1} - \theta_{2}$. In the
simulations, we found three types of stable orbits for HD
82943:(I)Only $\theta_{1}$ librates about $0^{\circ}$ (II)Case of
alignment, $\theta_{1}\approx\theta_{2}\approx\theta_{3}\approx
0^{\circ}$ (III)Case of antialignment, $\theta_{1}\approx
180^{\circ}$, $\theta_{2}\approx0^{\circ}$,
$\theta_{3}\approx180^{\circ}$. In addition, we noticed that 2\%,
3\% and 2\% of the cases belong to Classes I, II and III
respectively. The geometry of Class I was also discovered by
Gozdziewski \& Maciejewski (2001). But Classes II and III are
quite a new discovery that exhibits the HD 82943 system can
undertake apsidal lock (see Figure 3). Recently, Lee M. H.(2003,
private communication) pointed out that the best-fit
solution\footnote{The velocities for HD 82943 have not yet been
published, and the best-fit orbital parameters taken from
http://obswww.unige.ch.} makes the orbits closer to be aligned
than anti-aligned in the study of the 2:1 resonance. We used a
semi-analytical (Ji et al. 2003b) model to investigate the apsidal
motion for this system and found that although the two planets
have high eccentricities up to 0.54 and 0.41, the system remain
stable for 10 Myr because of both dynamical mechanisms of the 2:1
MMR and apsidal lock.

Several best-fit orbital solutions for GJ 876 appeared in
literature (Marcy et al. 2001; Laughlin \& Chambers 2001;Rivera \&
Lissauer 2001). Here we presented numerical results based on
Laughlin-Chambers fit. Two families of orbits are found to be
stable:Classes I (14\%) and II (7\%). In summary, for above two
systems, we stress that the 2:1 MMR always takes place for stable
cases and the apsidal alignment(Laughlin \& Chambers 2001;Lee \&
Peale 2002) seems to have more likely chances.

On the other hand, as the two orbits of GJ 876 (or HD 82943) are
not well separated, repeated interplanetary close approaches may
lead to disruption of the systems:79\% for GJ 876 rapidly become
unstable, and 93\% for HD 82943.

\subsection{Other systems}
Butler et al.(1999) reported the first triple-planet system of
$\upsilon$ And, consisting of a 51 Peg-like planet with a 4.6-day
orbit and two librating outer planets (Rivera \& Lissauer
2000;Chiang, Tabachnik \& Tremaine 2001;Chiang \& Murray 2002).
With the updated data(Paper II), we again examined several tests
for integration of $10^{6}$ yr and discovered that two outer
components can still remain in the apsidal lock about $0^{\circ}$.
As for 55 Cnc, we revealed that the inner two companions perform
the asymmetric $\Delta\varpi$-libration about $250^{\circ}$ or
$110^{\circ}$ in the secular evolution (Ji et al. 2003a).

Nagasawa et al.(2003) showed that the two planets of HD 168443 and
HD 74156 are both in circulation. Moreover,we performed additional
experiments for HD 38529, but simply found the planets circulate.
The systems of HD 168443, HD 74156 and HD 38529 are similar (Table
2), because each of them has a more massive outer planet and a
close inner companion. The reason for circulation will be
explained in \S 3. Nevertheless, if one of three systems can host
an extra middle planet with new observations, there may exist
librating planets similar to those of 55 Cnc and $\upsilon$ And.

\section{Discussion on apsidal motion of $\Delta\varpi$}
However, one should make clear the conditions for libration or
circulation of $\Delta\varpi$ for two planets. Several theoretical
works (Malhotra 2002;Lee \& Peale 2003;Nagasawa et al. 2003;Zhou
\& Sun 2003)were concentrated  on the problem. Beauge et al.(2003)
presented analytical and numerical outcomes on the existence and
location of stable equilibrium solutions about the systems in mean
motion resonance. Here we present our results based on L-L theory.
Following MD99, we have:
\begin{equation}
\label{eq1} \frac{d\varpi_{b}}{dt} = -
\frac{e_{c}}{e_{b}}A_{12}\cos (\varpi_{b} - \varpi_{c}) - A_{11}
\end{equation}
\begin{equation}
\label{eq2} \frac{d\varpi_{c}}{dt} = -
\frac{e_{b}}{e_{c}}A_{21}\cos (\varpi_{b} - \varpi_{c})  - A_{22}
\end{equation}
\begin{equation}
\label{eq3} \frac{d\Delta\varpi}{dt}=A_{22} - A_{11}+
(\frac{e_{b}}{e_{c}}A_{21}-\frac{e_{c}}{e_{b}}A_{12}) \cos
\Delta\varpi
\end{equation}
where $A_{11}$, $A_{12}$, $A_{21}$ and $A_{22}$ are initially
determined constants (see MD99) and $e_{b}$, $e_{c}$ are the
eccentricities of each planet. Here we rewrite the coefficients in
equation (4), such that $B_{1}=A_{22} - A_{11}$ and
$B_{2}=(e_{b}/e_{c})A_{21} - (e_{c}/e_{b})A_{12}$. For
hierarchical planetary systems (HD 168443, HD 74156 and HD 38529,
the ratio of the semi-major axes $\alpha=a_{b}/a_{c} \ll 1$), we
have $|B_{1}|\gg |B_{2}|$(see Table 2), thus the part of $|B_{1}|$
is dominant and $\Delta\varpi$ circulates(see also Nagasawa et
al.2003). According to (4), the critical values of $e_{b}/e_{c}$
for $\Delta\varpi$-libration:
\begin{equation}
({\frac{e_{b}}{e_{c}})}_{\pm}=\frac{\pm(\gamma-1)-
\sqrt{(\gamma-1)^{2}+4\gamma\beta^{2}}}{2\beta}
\end{equation}
where the plus sign in $\pm$ corresponds to the case for
$\Delta\varpi$-libration about $0^{\circ}$, and the minus sign to
that of $\Delta\varpi$-libration about $180^{\circ}$, with
$\beta=-{b^{(2)}_{3/2}(\alpha)}/{b^{(1)}_{3/2}(\alpha)}$
($b^{(j)}_{3/2}(\alpha)$ are Laplace coefficients, $j=1,2$) and
$\gamma=(1/{\sqrt\alpha})(\mu_{c}/\mu_{b})$$\sqrt{(1+\mu_{c})/(1+\mu_{b})}$
($\mu_{b,c}$ are respectively the ratio of the masses of Companion
b (c) and host star). Hence, there are two requirements for
apsidal libration of a pair of planets in the absence of mean
motion commensurabilities. Firstly, the ratio of initial
eccentricities of two planets should not be far from the critical
values;secondly, the starting relative apsidal longitude should be
satisfied to approach $0^{\circ}$ or $180^{\circ}$. As a paradigm
of HD 37124 (see Figure 1), the initial ratio of the
eccentricities $(e_{b}/e_{c})_{0}\simeq0.30$ is close to
$(e_{b}/e_{c})_{+}=0.34$, and
$\Delta\varpi_{0}\simeq7^{\circ}\simeq0^{\circ}$, thus the two
planets are aligned. We can apply the criteria to other systems.
\section{Summary}
In this work, we investigated the apsidal motion for the
multi-planet systems:we find that the two planets of HD 37124, HD
12661, 47 Uma and HD 82943 separately undergo the apsidal
alignment or antialignment. But the companions of GJ 876 and
$\upsilon$ And are only in apsidal lock about $0^{\circ}$, which
means that no mirror configurations occur for them. Moreover, we
again used the L-L theory to discern whether two planets of a
system librate or not. The status of $\Delta\varpi$-libration
depends on the ratio of their initial eccentricities,semi-major
axes and planetary masses and original relative apsidal
longitudes. And it seems to be a selection effect for the
companions in the multi-planet systems that are involved in the
apsidal libration. However, this should be further testified by
additional observations of the planetary systems.

\acknowledgments {We thank G. Laughlin, R.P. Butler, D.A. Fischer,
M.H. Lee for informative discussions and an anonymous referee for
good comments to improve the manuscript.This work is financially
supported by National Natural Science Foundation of China(Grants
10203005, 10173006, 10233020) and Foundation of Minor Planets of
Purple Mountain Observatory.}


\begin{deluxetable}{lccccccccc}
\tabletypesize{\footnotesize} \tablewidth{0pt} \tablecaption{The
astrocentric orbital parameters of seven multi-planet
systems\tablenotemark{1}} \tablehead{\colhead{Planet} &
\colhead{$M_{star}$ } & \colhead{$M\sin i$ } & \colhead{$P$ } &
\colhead{$a$ } &\colhead{$e$}
&\colhead{$\omega$ } &\colhead{}  &\colhead{Comments} &\colhead{}\\
\colhead{ } & \colhead{($M_{\odot}$)} & \colhead{($M_{Jup}$)} &
\colhead{(days)} & \colhead{(AU)} &\colhead{ } &\colhead{(deg)}
&\colhead{(MMR)} &\colhead{(Align)} &\colhead{(Anti)}
} \startdata
GJ  876  b\tablenotemark{a}   &0.32 &1.06  &29.995  &0.1294 &0.314(0.02)   &51.8(10)    &2:1  &Yes &No\\
GJ  876  c\tablenotemark{a}   &0.32 &3.39  &62.092  &0.2108 &0.051(0.02)   &40.0(10)    &2:1  &Yes &No\\
HD 82943 b\tablenotemark{b}   &1.05 &0.88  &221.6   &0.73   &0.54(0.05)    &138(13)     &2:1  &Yes &Yes\\
HD 82943 c\tablenotemark{b}   &1.05 &1.63  &444.6   &1.16   &0.41(0.08)    &96(7)       &2:1  &Yes &Yes\\
HD 37124 b\tablenotemark{c}   &0.91 &0.86  &153.0   &0.54   &0.10(0.06)    &97(40)      &     &Yes &Yes\\
HD 37124 c\tablenotemark{c}   &0.91 &1.01  &1942.0  &2.95   &0.40(0.20)    &265(120)    &     &Yes &Yes\\
47  UMa  b\tablenotemark{d}   &1.03 &2.86  &1079.2  &2.077  &0.05(0.04)    &124.3(12.7) &8:3? &Yes &Yes\\
47  UMa  c\tablenotemark{d}   &1.03 &1.09  &2845.0  &3.968  &0.00(0.03)    &170.8(15.3) &8:3? &Yes &Yes\\
HD 12661 b\tablenotemark{d}   &1.07 &2.30  &263.6   &0.823  &0.35(0.03)    &293.1(5.0)  &11:2/6:1 &Yes &Yes\\
HD 12661 c\tablenotemark{d}   &1.07 &1.57  &1444.5  &2.557  &0.20(0.04)    &162.4(18.5) &11:2/6:1 &Yes &Yes\\
Ups And  b\tablenotemark{d}   &1.30 &0.64  &4.617   &0.058  &0.019(0.20)   &200.0(20.0)                \\
Ups And  c\tablenotemark{d}   &1.30 &1.79  &241.16  &0.805  &0.26(0.03)    &249.1(9.3)  &16:3? &Yes &No\\
Ups And  d\tablenotemark{d}   &1.30 &3.53  &1276.15 &2.543  &0.25(0.03)    &264.2(17.2) &16:3? &Yes &No\\
55  Cnc  b\tablenotemark{e}   &1.03 &0.83  &14.65   &0.115  &0.03          &90.25       &3:1   &No  &No\\
55  Cnc  c\tablenotemark{e}   &1.03 &0.20  &44.27   &0.241  &0.41          &46.39       &3:1   &No  &No\\
55  Cnc  d\tablenotemark{e}   &1.03 &3.69  &4780.0  &5.461  &0.28          &218.91      &      &    &
\enddata

\tablenotetext{1} {Recently, Lee \& Peale (2003) pointed out the
difference between the astrocentric and Jacobi orbits for a
hierarchical system. The values are taken from the references:(a)
Laughlin et al.(2001) (b) http//obswww.unige.ch (c) Butler et al.
(2003) (d) Fischer  et al.(2003) (e) Fischer et al.(2003) and Ji
et al.(2003)}
\end{deluxetable}
\clearpage

\begin{deluxetable}{lccccc}
\tabletypesize{\small}
\tablewidth{0pt}
\tablecaption{The $\Delta\varpi$-circulation for three
hierarchical planetary systems}
\tablehead{\colhead{System\tablenotemark{a}}
&\colhead{$M_{b}/M_{c}$} &\colhead{$a_{b}/a_{c}$}
&\colhead{$e_{b}/e_{c}$} &\colhead{$B_{1}/B_{2}$}
&\colhead{($g_{+}/g_{-}$)\tablenotemark{b}}}
\startdata
HD 38529       &0.06     &0.035    &0.81   &18.3    &87.3 \\
HD 74156       &$<$0.20  &$<$0.072 &1.86   &23.6    &$>$19.0 \\
HD  168443     &0.45     &0.1027   &2.65   &7026.8  &7.0
\enddata
\tablenotetext{a}{The data for the masses, eccentricities and
semi-major axes of  two planets for three systems are taken from
Fischer et al.(2003).}
\tablenotetext{b} {Here $g_{+}$, $g_{-}$
are two eigenfrequencies from L-L secular solutions. For
$\alpha=a_{b}/a_{c}\ll 1$, we have $g_{+}/g_{-}=(1/{\sqrt\alpha})
(\mu_{c}/\mu_{b})\sqrt{(1+\mu_{c})/(1+\mu_{b})}=\gamma$. And
$g_{+}/g_{-} \simeq (1/{\sqrt\alpha})(\mu_{c}/\mu_{b})\gg 1 $,
indicating $g_{+}$ is dominant.}
\end{deluxetable}

\clearpage
\epsscale{0.90}
\begin{figure}
\figurenum{1} \plotone{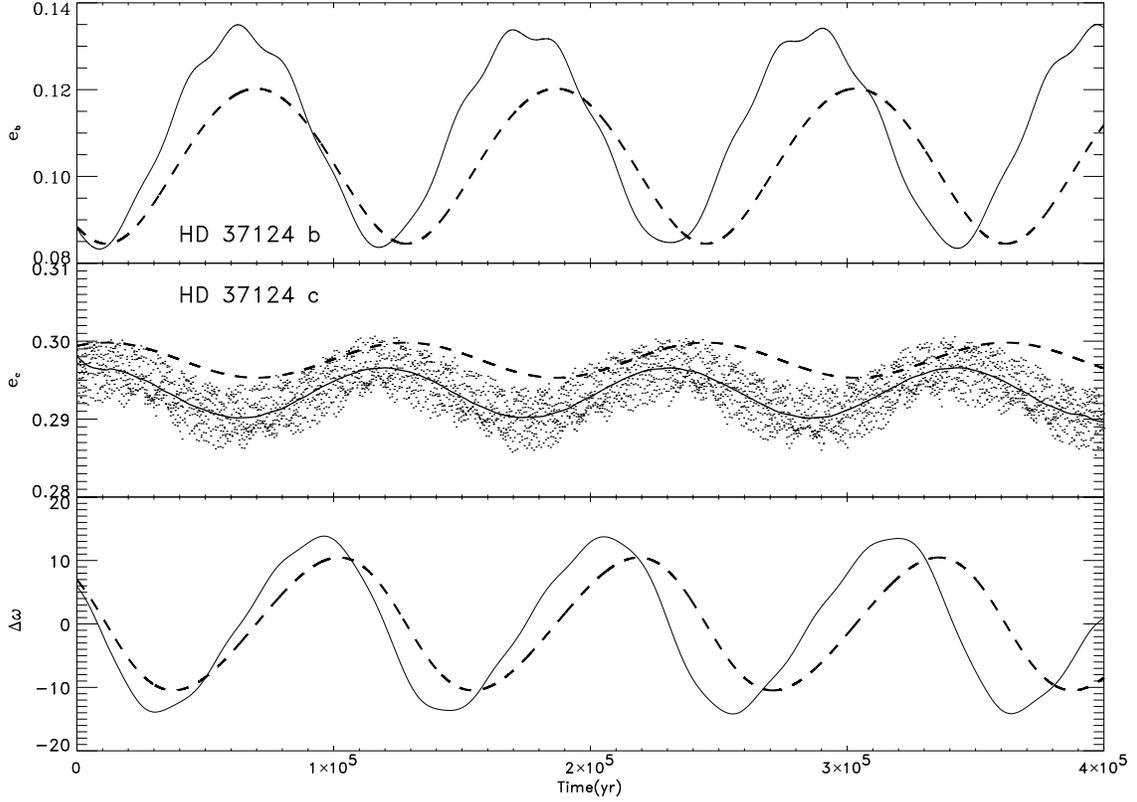} \caption{The alignment case for HD
37124. The upper two panels show the secular evolution of the
eccentricities of Companions b and c. Notice that $e_{b}$ ranges
from 0.08 to 0.14, while $e_{c}$ varies from 0.28 to 0.30 (Case
b). The bottom panel displays the relative apsidal longitude
$\Delta\varpi$ librates about $0^{\circ}$ with the amplitude of
$\pm15^{\circ}$. The thick lines denote the numerical results
(dots for $e_{c}$) and the dash lines are those from L-L theory.
Both of them are in good agreement.The initial values:$a_{b}=0.54$
AU,$e_{b}=0.089$,$\Omega_{b}=199.22^{\circ}$,$\omega_{b}=83.38^{\circ}$.
$a_{c}=2.95$AU, $e_{c}=0.299$,$\Omega_{c}=344.54^{\circ}$,
$\omega_{c}=291.10^{\circ}$. \label{fig1}}
\end{figure}

\clearpage
\begin{figure}
\figurenum{2} \plotone{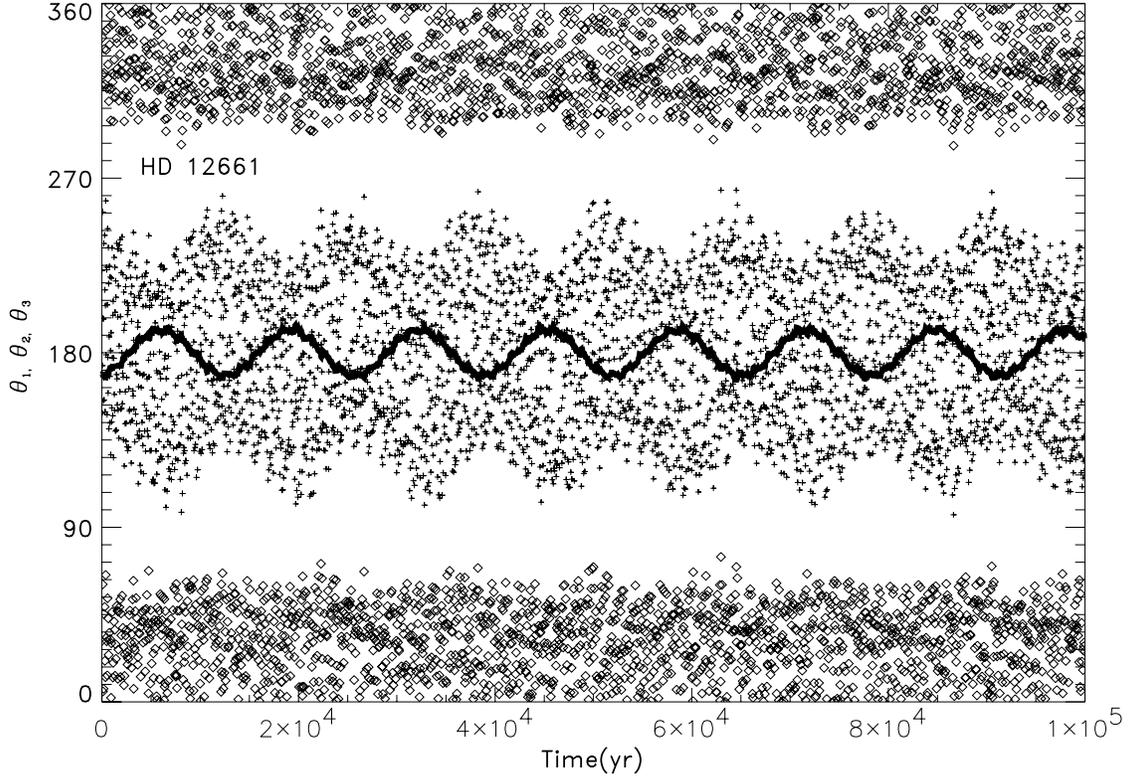} \caption{The antialignment case for
HD 12661.  Here $\theta_{1} = 2\lambda _{b} - 11\lambda _{c} +
3\varpi_{b} + 6\varpi_{c}$, $\theta_{2} = 2\lambda _{b} -
11\lambda _{c} + 2\varpi_{b} + 7\varpi_{c}$ and $\theta_{3} =
\Delta\varpi$. The signs of diamond and plus, respectively,
represent $\theta_{1}$ and $\theta_{2}$. And the thick line
denotes $\theta_{3}$, which librates about $180^{\circ}$ with
small amplitudes for $10^5$ yr. Notice that $\theta_{1}$ librates
about $0^{\circ}$, while $\theta_{2}$ does about $180^{\circ}$ for
the same timescale. In fact, ten resonant arguments $\phi_{k}$
($k=0...9$) all librate for the same timescale for this case. The
scenario can be seen for 1 Myr.\label{fig2}}
\end{figure}

\clearpage
\begin{figure}
\figurenum{3} \plotone{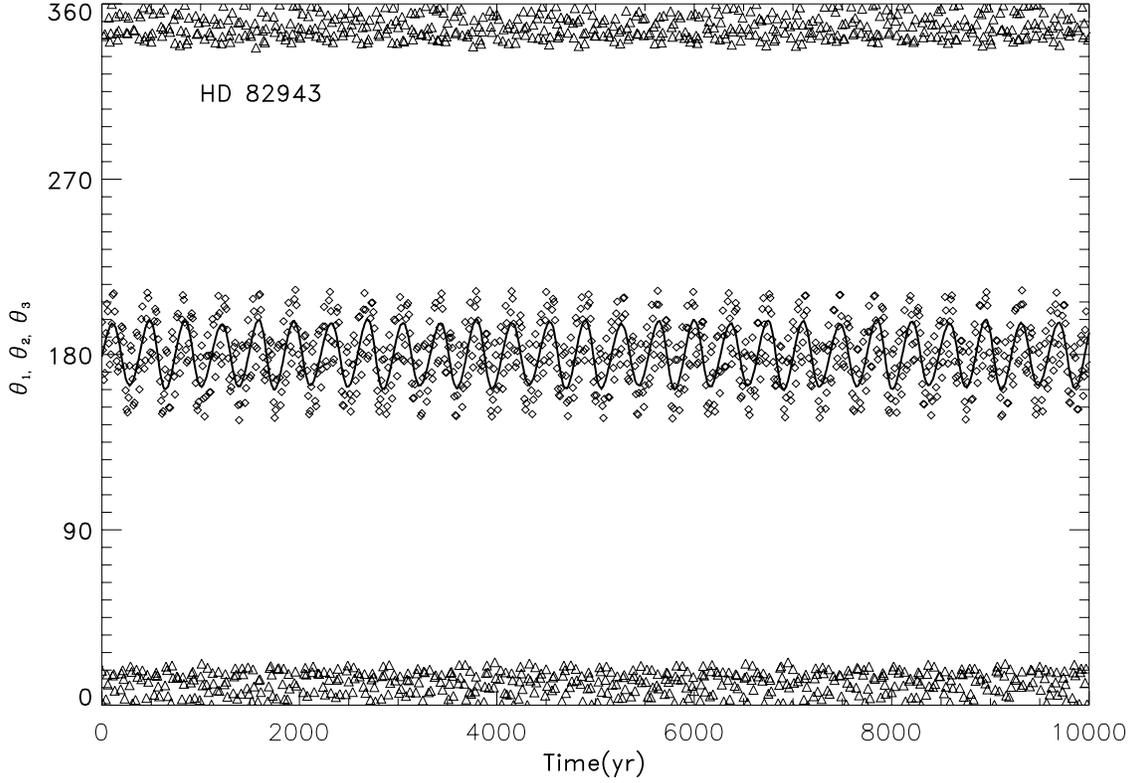} \caption{The antialignment case for
HD 82943 (Class III). Here $\theta_{1} = \lambda _{b} - 2\lambda
_{c} + \varpi_{b}$, $\theta_{2} = \lambda _{b} - 2\lambda _{c} +
\varpi_{c}$ and $\theta_{3} = \Delta\varpi$. Note that two
critical arguments of $\theta_{1}$ (the sign triangle) and
$\theta_{2}$ (the sign diamond), respectively, librate about
$0^{\circ}$ and $180^{\circ}$ with slight amplitudes for $10^4$
yr. The thick line represent $\theta_{3}$, librating about
$180^{\circ}$ in the same timescale. The results can extend to 10
Myr. \label{fig3} }
\end{figure}

\end{document}